\documentclass[intlimits,twoside,a4paper]{article}

\usepackage{amsmath,amssymb}
\usepackage{graphicx}

\usepackage[T2A]{fontenc}
\usepackage[cp1251]{inputenc}
\usepackage{color}

\usepackage[eqsecnum]{cmpj2}

\newcommand{\ang}[1]{{\langle #1\rangle}}
\newcommand{\lp}{\left(}
\newcommand{\rp}{\right)}
\newcommand{\ls}{\left[}
\newcommand{\rs}{\right]}

\newcommand{\be}{\begin{equation}}
\newcommand{\ee}{\end{equation}}
\newcommand{\suml}{\sum\limits}
\newcommand{\intl}{\int\limits}
\newcommand{\ptsed}[1]{\left(#1\right)}

\issue{2013}{16}{3}{33705}
\doinumber{10.5488/CMP.16.33705}

\title[Magnetic states of single impurity in disordered environment]{Magnetic states of single impurity \\ in disordered environment
}
\author[G.V. Ponedilok, M.I. Klapchuk]{G.V. Ponedilok, M.I. Klapchuk}
\address{National University ``Lviv Polytechnic'',
Institute of Applied Mathematics and Fundamental Sciences, \\
12 S.~Bandera St., 79013 Lviv, Ukraine}

\date{Received April 12, 2013, in final form June 18, 2013}
\authorcopyright{G.V. Ponedilok, M.I. Klapchuk, 2013}

\begin{document}

\maketitle

\begin{abstract}
The charged and magnetic states of isolated impurities dissolved in
amorphous metallic alloy are investigated.
The Hamiltonian of the system under study
is the generalization of Anderson impurity model. Namely,
the processes of elastic and non-elastic scattering of conductive
electrons on the ions of a metal and on a charged impurity
are included.
The configuration averaged one-particle Green's functions are
obtained within Hartree-Fock approximation. A system of
self-consistent equations is given for calculation of an electronic
spectrum, the charged and the spin-polarized impurity states. Qualitative analysis of the effect of the metallic host
structural disorder on the observed values is performed. Additional shift
and broadening of virtual impurity level is caused by a structural
disorder of impurity environment.
\pacs 71.23. -k, 71.23. An, 72.15. Rn
\end{abstract}

\section{Introduction}

The purpose of this work is to explore the effects of the structural
disorder on the states of electronegative impurities
dissolved in liquid alkali metal.
The ions belonging to metal matrix form a
complicated random field for an impurity atom. The model proposed in this study
is applicable to a structurally disordered system in which the
tight-binding representation of the electronic wave function is
appropriate and the effect of a short-range order is eminent.
The liquid alkali metals as well as  amorphous solids are the
systems we would like to investigate.

In this work we describe the states of isolated impurities of such
elements as H, O, Cl, F, N using generalized microscopic theory
based on the single-impurity Anderson model (SIAM) \cite{Anderson}.
While discussing the macroscopic features of the single-electron
properties of the system, the procedure of effective Green's function
ensemble-averaging over all possible configurations of atoms
${\bf R}_1,\ldots, {\bf R}_N$ is necessary.
The procedure of configurational averaging is
an enormously difficult problem in the multiple-scattering theory.
Only the two-particle correlation functions are known from
experimental data.
In practice, of course, our knowledge of these density correlation
functions is incomplete and various approximate theories for
short-range order involve only the one- and two-site distribution
functions \cite{edwards,roth,yonezawa,kanejoshi,shwartz}.
The short-range order is always present in liquid metals; its simplest
manifestation is in the characteristic oscillation of the x-ray
structure factor or in the oscillation of radial distribution function.

The main goal of this paper is to show the effect of disordered
local impurity environment on its charge and magnetic states.
The experimentally observed magnetic moment decrease for some
sorts of ferromagnetic solids or amorphous alloys
is discussed in detail in~\cite{xandrix}.

Microscopic model to describe the electronegative impurity in a disordered
system is proposed in section~1. The Hamiltonian of the system is a generalization
of Anderson impurity model. It also includes the processes of
elastic and non-elastic scattering of conductive electrons on the ions of a
metal and on a charged impurity. Qualitative and quantitative
estimates of the parameters of the Hamiltonian have been carried out in
\cite{prepr0213}. The formation of an effective charge and
spin-polarized gaseous impurity states in a liquid metal can be
described as the process of hybridization of local level with
quasi-free electron states under the effect of a polarizing impurity
potential~\cite{prepr0224}.

The two-time retarded Green's functions \cite{Zubarev} are obtained
within Hartree-Fock (HFA) approximation in section~3. The configuration
averaged system of Green's functions is obtained in section~4. A system
of self-consistent equations is given for the calculation of the electronic
spectrum, as well as the charged and the spin-polarized impurity states.
The qualitative analysis of the effect of the metallic host
structural disorder on the observed values is performed. An additional shift
and broadening of a virtual impurity level is caused by the structural
disorder of an impurity environment.

\section{Microscopic model of the system}

Let us consider a  single impurity dissolved in liquid
alkaline metal. The liquid metal phase will be described within the
framework of electron-ion model which for such metals gives
satisfactory computational results for electronic and structural
properties.
Let ${\bf R}_1,\dots,{\bf R}_N$ be the coordinates of atoms of
metallic alloy which take arbitrary values in the volume $V $. The
impurity has a coordinate $ {\bf R}_0 $. We have chosen the following
model Hamiltonian in coordinate representation:
\be
\hat H = H_\textrm{cl}+\hat H_\textrm{el-i}+\hat H_\textrm{el-el}\,.
\label{H}
\ee
The energy operator of electron-ion interaction is
written as follows:
\be
\hat H_{\rm el-i}=-\frac{\hbar^2}{2m}\sum_{1\leqslant i\leqslant N}\Delta_i +
\sum_{1\leqslant i\leqslant N}\sum_{1\leqslant j\leqslant N}V(\mid{\bf r}_{i}-{\bf
R}_{j}\mid)
+\sum_{1\leqslant i\leqslant N_e} V_{0}(\mid{\bf r}_{i}-{\bf R}_0\mid).
\label{H-el-i}
\ee
In this equation $ {\bf r}_1, \dots, {\bf r}_ {N} $ are the electron
coordinates of a metallic subsystem, the amount of which coincides
with the number of metal atoms due to single valence of alkaline
elements. It is assumed that the electrons of valence impurity
orbital remain localized on the impurity.

The potentials $V(|{\bf r}_i-{\bf R}_j|)$ and $V_0(|{\bf r}_i-{\bf
R}_0|)$ describe electron scattering on ions of metal and impurity,
respectively. The first term in equation (\ref{H-el-i}) is the operator of
a kinetic energy of free electron subsystem.

The last term in (\ref{H}) describes the energy of pair
electron-electron interaction
\be
\hat H_\textrm{el-el}{=}\frac{1}{2}\sum_{1\leqslant i\ne j\leqslant N}\Phi(\mid
{\bf r}_{i}-{\bf r}_{j}\mid){=}\frac{1}{2}\sum_{1\leqslant i\ne j\leqslant
N}\frac{e^2}{\mid{\bf r}_{i}-{\bf r}_{j}\mid}\,.
\label{H-el-el}
\ee
The non-operator part, $H_{\rm cl}$, describes the energy of classical
ion-ion interaction.

In order to represent the secondary quantization, we
use plane waves as a basis in order to decompose the field electronic operators
\be
\varphi_{\bf k}({\bf r})=\frac1{\sqrt V}\exp{(\ri{\bf k}\cdot {\bf r})}
\label{varphi}
\ee
and s-shell localized on the impurity
\be
\psi_{0}({\bf r})=\sqrt{\frac1{\pi r_\textrm{p}^{3}}}\exp{\ptsed{-\frac{|{\bf
r-R}_0|}{r_\textrm{p}}}}.
\label{psi}
\ee

The wave vector ${\bf k}$ in (\ref{varphi}) takes specified
values in the impulse quasi-continuous space $\Lambda$:
\be
\Lambda=\,\bigg\{\,{\bf k}:\,{\bf k}=\sum_{1\leqslant
\alpha\leqslant3}2\pi\,V^{1/3}n_{\alpha}{\bf e}_{\alpha},\
n_{\alpha}\in{Z} ,\ ({\bf e}_{\alpha},{\bf
e}_{\beta})={\delta}_{\alpha\beta}\bigg\}.
\ee

Let us mention that $\psi_{0}({\bf r})$ is not orthogonal to the plane
waves (\ref{varphi}). Apart from this, its inclusion into the basis set
causes overfilling of the latter. However, the inaccuracy introduced by
such an approximate procedure will not affect the qualitative picture.
In the  representation of the secondary
quantization operator (\ref {H}) with allowance for only a certain
class of Coulomb electron-electron interactions we then have the
following expression,
\begin{eqnarray}
\hat H&=&H_{\rm cl}+\suml_{{\bf
k}\in\Lambda}\suml_{\sigma=\pm1}\,E_k\,a_{\bf k\sigma}^+ \,a_{\bf
k\sigma}+\suml_{\sigma=\pm1} E_0\,d_{0\sigma}^+\,d_{0\sigma}
+\suml_{{\bf k}\in\Lambda}\suml_{{\bf
q}\in\Lambda}\suml_{\sigma=\pm1}\ptsed{V_{\bf q}\, a_{\bf
k\sigma}^+\,a_{\bf k-q,\sigma}+V_{0,{\bf q}}\,a_{\bf k\sigma}^+\,a_{\bf
k-q,\sigma}}\nonumber\\
%\]
%\[
&&+\suml_{\sigma=\pm1}U_0\,\hat n_{0\sigma}\hat
n_{0,-\sigma}+\suml_{{\bf k}\in\Lambda}\suml_{\sigma=\pm1}\ptsed{W_{{\bf k},0}\,a_{\bf
k\sigma}^+\,d_{0\sigma}+W_{{\bf k},0}^*\,d_{0\sigma}^+\,a_{\bf k\sigma}}
+\suml_{{\bf k}\in\Lambda}\suml_{{\bf q}\in\Lambda}\suml_{\sigma,
\sigma'=\pm1}P_{{\bf q},0}\,a_{\bf k\sigma}^+\,a_{\bf k-q,\sigma}\,\hat
n_{\sigma'}\nonumber\\
%\]
%\be
&&+\suml_{{\bf
k}\in\Lambda}\suml_{\sigma\neq\sigma'}\ptsed{U_{{\bf k},0}\,\hat
n_{\sigma'}\,a_{\bf
k\sigma}^+\,d_{0,\sigma}+U_{{\bf k},0}^*\,d_{0,\sigma}^+\,a_{\bf
k\sigma}\,\hat n_{\sigma'}}.
\label{Hamiltonian}
\end{eqnarray}
Here, $a_{\bf k\sigma}(a_{\bf k\sigma}^+)$ and $d_{0,\sigma}(d_{0,\sigma}^+)$ are
the annihilation (creation) Fermi-type operators for electrons in the
states  $\{{\bf k},\sigma\}$ and $\{{\bf R}_0,\sigma\}$, where
$\sigma{=}\pm1$ is quantum spin number, which takes two values
due to two possible orientations of electronic spin
relatively to the quantization axis. $E_k=\hbar^2k^2/2m$ is the energy
spectrum of the electrons in states $\varphi_{\bf k}({\bf r})$, and
$E_0$ is the energy of the localized electronic state $\psi_0({\bf
r})$. $\hat n_\sigma=d_\sigma^+\,d_\sigma$ is the spin-dependent
occupation number operator for the localized state.

The matrix elements  $V_{\bf q}$ and $V_{0,{\bf q}}$ characterize the processes
of elastic scattering of electrons on the ions of the metal and on
the impurity. Their explicit analytical forms are as follows:
\be
V_{\bf q}=\frac1N\suml_{1\leqslant j\leqslant N}\,\re^{-\ri{\bf q}\cdot{\bf R}_j}\,v(q),\qquad
V_{0,{\bf q}}=\re^{-\ri{\bf q}\cdot{\bf R}_0}\,v_0(q).
\ee
The formfactors of the scattering potentials
\be
v(q)=\intl_VV(|{\bf r}|)\,\re^{-\ri{\bf q}\cdot{\bf r}}\,\rd{\bf
r}, \qquad v_0(q)=\frac1V\intl_VV_0(|{\bf r}|)\,\re^{-\ri{\bf q}\cdot{\bf r}}\,\rd{\bf
r}
\label{formfactors}
\ee
depend only on the absolute value of the momentum transfer $ {\bf q} $
due to the locality of the potentials $V(|{\bf r}|) $ and $V_0
(|{\bf r}|) $.

The processes of inelastic scattering of electrons caused by their
transition from the state localized on the impurity into the conduction
band and vice versa are characterized
 by the matrix element,
\be
W_{{\bf k},0}=\frac1{\sqrt{V}}\intl_V \re^{-\ri{\bf
k}\cdot{\bf r}}\left[-\frac{\hbar^2\Delta_{\bf r }}{2m}+V_{\rm LF}({\bf
r})\right]\,\psi_0({\bf r})\,\rd{\bf r}.
\label{W-k}
\ee
Here,
\be
V_{\rm
LF}({\bf r})=\suml_{1\leqslant j\leqslant N}V(|{\bf r-\bf R}_j|)+V_0(|{\bf
r-\bf R}_0|)
 \label{V-LF}
\ee
is the potential of a local field of metal ions and the
impurity, which acts on the electron at a point $ {\bf r} \in V $.

The term  $\sum_{\sigma}U_0\,\hat n_\sigma\hat n_{-\sigma}$ in the
Hamiltonian (\ref{Hamiltonian}) arises from the  operator
of Coulomb  electron interaction and describes the Hubbard repulsion
of electrons localized on the impurity with the intensity  $U_0$.
\be
U_0=\int \rd{\bf r}_1\int \rd{\bf r}_2|\psi_0({\bf r}_1)|^2
\frac{e^2}{|{\bf r}_1-{\bf r}_2|}|\psi_0({\bf r}_2)|^2=
\frac{5}{8}\frac{e^2}{r_\textrm{p}}\,,
\ee
this value is approximately about $1\div5$~eV for the atom of oxygen.

The process of elastic and inelastic scattering of electrons on the charged
impurity is described by the matrix elements,
\be
 P_{{\bf q},0}=\intl_V\,\re^{-\ri{\bf q}\cdot{\bf r}}\,\widetilde\Phi({\bf r})\,\rd{\bf r},\qquad
 U_{{\bf k},0}=\frac1{\sqrt{V}}\intl_V \re^{-\ri{\bf
k}\cdot{\bf r}}\widetilde\Phi({\bf r})\psi_0({\bf r})\,\rd{\bf r}.
\label{P-q}
\ee
Here, the value
\be
\widetilde\Phi({\bf
r})=\intl_V\,\Phi(|{\bf r-r}^\prime|)\,|\psi_0({\bf r}^\prime)|^2 \,
\rd{\bf r}^\prime,
\label{Phi}
\ee
gives the potential energy
of the electron in a field which is generated by the electron
localized on the  $\psi_0 ({\bf r})$ orbital.

The matrix elements (\ref{W-k})--(\ref{P-q}) can be written down in the other form
by separating explicitly the structural multipliers
\be
W_{{\bf k},0}=\re^{-\ri{\bf
k}\cdot{\bf R}_0}\,w_k\, ,
\qquad
U_{{\bf k},0}=\re^{-\ri{\bf k}\cdot{\bf R}_0}\,u_k\, ,
\qquad
P_{{\bf k},0}=\re^{-\ri{\bf k}\cdot{\bf R}_0}\,p_k\,.
\label{1.13}
\ee
The coefficients $w_k,u_k,p_k$ do not depend here on the nodal index
and are considered in the coordinate system related to the impurity.
Their analytical form is given in~\cite{JPS}.

Actually, only the electrostatic
effects including two electrons are taken into account
in the Hamiltonian (\ref{Hamiltonian}),
while  the processes of exchange are not considered.

%%%%%%%%%%%%%%%%%%%%%%%%
%%%%%%%%%%%%%%%%%%%%%%%

\section{Green's function method. Hartree-Fock approximation}

The method of equation of motion of Green's functions is one of the most important
tools to solve the model Hamiltonian problems in condensed-matter physics~\cite{Hong}.
Let us calculate the matrix of retarded time-dependent temperature Green's functions
\be
{\bf  G}(\omega)=\begin{pmatrix}
G_{{\bf k},{\bf k}'}^\sigma(\omega)  &  M_{{\bf k},0}^\sigma(\omega)\cr
M_{0,{\bf k}'}^\sigma(\omega)        &  L_{0,0}^\sigma(\omega)\cr\end{pmatrix}
 \equiv\begin{pmatrix} \ang{\ang{a_{{\bf k}\sigma}|a_{{\bf
k}'\sigma}^+}}_\omega & \ang{\ang{a_{{\bf
k}\sigma}|d_{0\sigma}^+}}_\omega\cr \ang{\ang{d_{0\sigma}|a_{{\bf
k}'\sigma}^+}}_\omega &
\ang{\ang{d_{0\sigma}|d_{0\sigma}^+}}_\omega\cr\end{pmatrix}.
 \label{G}
\ee
The equation of motion for each component of (\ref{G}) is given in our
earlier work~\cite{prepr0213,prepr0224,modpsepot,JPS}.
We make use of the decoupling scheme that corresponds to the Hartree-Fock  approximation type
for higher order Green functions.
The limits of the HFA applicability for the
description of real systems are considered in~\cite{Hong,Hubb,Hewson,HaldAnder,Haldane,Kishore}.

A set of connected Green's functions is obtained
\begin{eqnarray}
&&(\omega-E_k)G_{{\bf k},{\bf q}}^\sigma(\omega) =
\delta_{{\bf k},{\bf q}}+\sum\limits_{\bf p}\Lambda_{{\bf k}-{\bf p}}
G_{{\bf p},{\bf q}}^\sigma(\omega)
+\Omega_k^\sigma M _{0,{\bf q}}^\sigma(\omega),\label{3.12}\\
%\ee
%\be
&&(\omega-E_{0,\sigma})M_{0,{\bf q}}^\sigma(\omega)=
\sum\limits_{\bf p}{\Omega_\textrm{p}^*}^\sigma
G_{{\bf p},{\bf q}}^\sigma(\omega),
\label{3.13} \\
%\ee
%\be
&&(\omega-E_k)M_{{\bf k},0}^\sigma(\omega)=
\sum\limits_{\bf q}\Lambda_{{\bf k}-{\bf
q}}M_{{\bf q},0}^\sigma(\omega)
+\Omega_k^\sigma
L_{0,0}^\sigma(\omega),\label{3.14}\\
%\ee
%\be
&&(\omega-E_{0,\sigma})L_{0,0}^\sigma(\omega)=
1+\sum\limits_{\bf k}{\Omega_k^*}^\sigma
M_{{\bf k},0}^\sigma(\omega).
\label{3.15}
\end{eqnarray}
In the equations (\ref{3.12})--(\ref{3.15}), we use the notation
\begin{eqnarray}
\Lambda_{\bf q}=\sum\limits_{\alpha=1}^2V_{\bf q}^{(\alpha)}=
\frac{1}{N}\sum\limits_{1\leqslant j\leqslant N}\re^{-\ri{\bf q}\cdot{\bf R}_j}
v(|{\bf q}|)+\widetilde v_0(|{\bf q}|).
\label{3.16}
\end{eqnarray}

The Fourier-component of an effective impurity potential,
\be
\widetilde v_0(q)=v_0(q)+p_q\ang{\widehat n_0}=
\frac{1}{V}
\int\limits_{V} \re^{-\ri{\bf q}\cdot{\bf r}}\bigg[V_0(r)
+\ang{\widehat n_{0}}\int\limits_{V} \rd{\bf r}'\,|\psi_0({\bf r}')|^2\,
\Phi(|{\bf r}-{\bf r}'|)\bigg]\,\rd{\bf r}
\label{pseudopot}
\ee
\noindent
includes the Hartree-Fock potential, caused by the impurity atom
$\ang{\widehat n_{0}}=\sum_{\sigma}\ang{\widehat n_{0\sigma}}$.

In close similarity, the matrix elements  $\Omega_q^\sigma$ can be represented in the form,
$\Omega_q^\sigma=[u_q\ang{\widehat n_{-\sigma}}+w_q]$, or
\be
\Omega_q^\sigma=\frac{1}{\sqrt V}\int\limits_{V} \re^{-\ri{\bf q}\cdot{\bf r}}
\bigg[-\frac{\hbar^2{\pmb\nabla}^2}{2m}+\widetilde V_{\rm LF}^\sigma({\bf
r})\bigg]\psi_0({\bf r})\,\rd{\bf r},
\label{3.19}
\ee
where
\[
\widetilde V_{\rm LF}^\sigma({\bf r})=V_{\rm LF}({\bf r})+
\ang{\widehat n_{0,-\sigma}}\int\limits_{V} \rd{\bf r}'|\psi_0({\bf r}')|^2\Phi(|{\bf
r}-{\bf r}'|).
\]

From the equations (\ref{3.13}), (\ref{3.15}) one can find locator Green's function
\be
{L_{0,0}^\sigma(\omega)}=\frac{1}{ \omega-E_{0,\sigma}
-\sum\limits_{{\bf k,\,\bf q}} \Omega_k^\sigma \Lambda_{\bf
k,\bf q}(\omega)\,{\Omega_q^*}^\sigma}\,,
\label{L_loc}
\ee
here, we introduced the effective potential $\Lambda_{\bf k,\bf q}$, which
has the form of a series in terms of $\Lambda_{\bf
k-q}$
\be
\Lambda_{\bf k,\bf q}(\omega){=}\frac{\delta_{\bf k,
q}}{\omega{-}E_k{-}\Lambda_0}{+}\frac{\Lambda_{\bf
k-q}}{(\omega{-}E_k{-}\Lambda_0)(\omega{-}E_q{-}\Lambda_0)}+
\sum\limits_{{\bf p}} \frac{\Lambda_{\bf k-p}\Lambda_{\bf p-q}}
{(\omega{-}E_k{-}\Lambda_0)(\omega{-}E_\textrm{p}{-}\Lambda_0)(\omega{-}E_q{-}\Lambda_0)}
+\cdots \, .
\label{3.28}
\ee
Note that $\Lambda_0=v(0)+v_0(0)-2\pi\ang{\hat
n_0}e^2r_\textrm{p}^2/V.$

The non-diagonal Green function  $M_{0,\bf k}^\sigma(\omega)$, $M_{{\bf
k},0}^\sigma(\omega)$  and the propagator $G_{\bf
k,k'}^\sigma(\omega)$ are expressed by the locator
 $L_{0,0}^\sigma(\omega)$:
\begin{eqnarray*}
&&M_{0,\bf k}^\sigma(\omega){=}\sum\limits_{\bf
q}L_{0,0}^\sigma(\omega){\Omega_q^*}^\sigma \Lambda_{\bf
q,k}(\omega),\\
%\]
%\[
&&M_{{\bf k},0}^\sigma(\omega){=}[M_{0,\bf
k}^\sigma(\omega)]^*,\\[1ex]
%\]
%\[
&&G_{\bf k,k'}^\sigma(\omega){=}\Lambda_{\bf k,
k'}(\omega){+}\sum\limits_{\bf q,p}\Lambda_{\bf k,
q}(\omega)\Omega_q^\sigma
L_{0,0}^\sigma(\omega){\Omega_\textrm{p}^*}^\sigma \Lambda_{\bf p,
k'}(\omega).
%\label{Big_G}
\end{eqnarray*}

The renormalized impurity level is
\be
E_{0,\sigma}=E_0+U_0\,\ang{\widehat n_{0,-\sigma}}+
\sum\limits_{\bf k}\ls U_{{\bf k},0}\,\ang{a_{\bf
k,-\sigma}^+d_{0,-\sigma}}+
U_{{\bf k},0}^*\,\ang{d_{0,-\sigma}^+a_{\bf k,-\sigma}}\rs+
\sum\limits_{\bf k,\bf q}\sum\limits_{\sigma'}\,P_{{\bf q},0}\,
\ang{a_{\bf k,\sigma'}^+a_{\bf k-\bf q,\sigma'}}.
\label{3.21}
\ee
%%%%%%%%

%%%%%%%%%%%%%%%%%%%%%%%%%%%%%%%%%%%%%%%%%%%%%%%%%%%%%%%%%%%%%%%%%%%%%%%%
%%%%%%%%%%%%%%%%%%%%%%%%%%%%%%%%%%%%%%%%%%%%%%%%%%%%%%%%%%%%%%%%%%%%%%%%
\section{Configuration averaged Green's function}

One can start the averaging over all atomic configurations from equation~(\ref{L_loc}).
Here, the self-energy part describes the quasi-particles correlation.
The problem is specified in terms of (\ref{3.28}) describing the degree of correlation.
As the first approximation, we take into account  $\Lambda_{\bf k-q}$ only, while the higher order
correlation functions are neglected.

As common, we use a notation for the Fourier-transform of the atomic density fluctuations,
\[
\rho_{\bf k}=\frac{1}{\sqrt N}\sum\limits_{1\leqslant j\leqslant N}\re^{-\ri{\bf
k}\cdots{\bf R}_j},\qquad {\bf k}\ne0,
\]
\[
\Lambda_{\bf k-q}=\rho_{\bf k-q}v({\bf|k-q|})+\widetilde v_0({\bf|k-q|}).
\]

The configuration averaged Green's function of localized electrons is given as
\be
\overline{L_{0,0}^\sigma(E)}=\frac{1}{E{-}E_\sigma{-}\Sigma_0(k)}
\left\{1{+}\suml_{\frac{\bf k,q}{(\bf k\ne\bf
q)}}\frac{\Omega_k^\sigma{(\Omega_q^\sigma)}^*
\overline{\Lambda_{\bf
k-q}}}{(E{-}E_k{-}\Lambda_0)(E{-}E_q{-}\Lambda_0)[E{-}E_\sigma{-}\Sigma_0(k)]}{+}\cdots
\right\}\, ,
\ee
where
\[
\Sigma_0(k)=\sum\limits_{\bf
k}\frac{|\Omega_k^\sigma|^2}{E-E_k-\Lambda_0}
\]
is the self-energy term in the quasi-crystalline approximation.

We would like to discuss the case of inhomogeneous environment, where the impurity atom
gives origin to the spherically symmetrical potential. Therefore, $\overline{\rho_{\bf k-q}}=n_{\bf k-q}$,
in contrast to homogeneous case. Here,
$n({\bf r})=(1/V)\sum_{\bf k}n_{\bf k}\re^{\ri{\bf k}\cdot{\bf r}}$.
Taking into account the binary distribution function $n(r)$,
the expression for the averaged locator Green function is obtained:
\begin{equation}
\overline{L_{0,0}^\sigma(\omega)}= \left\{\omega-E_{0,\sigma}-\Sigma_0(k)-
\sum\limits_{{\bf k,\,\bf q}\atop{(\bf k\ne\bf q)}}
\frac{\Omega_k^\sigma\Omega_q^\sigma\ls n_{\bf
k-q}v({\bf |k-q|})+\,\widetilde v_0({\bf
|k-q|})\rs}{(E-E_k-\Lambda_0)(E-E_q-\Lambda_0)}\right\}^{-1}.
\label{locator}
\end{equation}

In order to get the density of states per atom for localized  electrons,  $\rho_0^\sigma(E)$ with spin $\sigma$,
we need to calculate the sum over $\bf k$ in (\ref{locator}).
\begin{eqnarray}
\lim_{\varepsilon\rightarrow
0}\sum\limits_{\bf k}\frac{|\Omega_k|^2}{
E-E_k-\Lambda_0+\ri\varepsilon}= {\cal
P}\sum\limits_{\bf k}\frac{|\Omega_k|^2}{ E-E_k-\Lambda_0}-\ri\pi\sum\limits_{\bf k}|\Omega_k|^2\delta(E-E_k-\Lambda_0),
\label{sumappr}
\end{eqnarray}
For the sake of convenience let us introduce the notation,
\begin{eqnarray}
&&\Delta^\sigma(E)=\pi\sum\limits_{\bf k}|\Omega_k|^2
\delta(E{-}E_k{-}\Lambda_0);\\
%\ee
%\be
&&\Lambda^\sigma(E)={\cal
P}\sum\limits_{\bf k}\frac{|\Omega_k|^2}{(E-E_k-\Lambda_0)}=
\frac1\pi{\cal P} \int \rd\,E^\prime
\frac{\Delta(E^\prime)}{E-E^\prime}\,.
\end{eqnarray}
The scattering of the s-electrons and localized electrons cause the impurity level to shift and
become broader. Namely,
 $\Lambda(E)$ is the effective shift whereas $\Delta(E)$ is the effective broadening of impurity level.
For the sake of simplicity, the matrix elements $(\Omega^\sigma)^2$ are estimated at the
Fermi level \cite{Hewson}.
Then, $\Delta^\sigma(E)$, $\Lambda^\sigma(E)$ are slowly varying functions
of $E$ over the band, and they can be treated as parameters,
\begin{align}
\Delta^\sigma(E)&=\pi{(\Omega^\sigma)}^2 \rho_0(E),\\
%\ee
%\be
\Lambda^\sigma(E)&={(\Omega^\sigma)}^2 \rho_0(E) g(E).
\end{align}
Here,
$$
\rho_0(E)=\frac{m^{3/2}}{2\sqrt{2}\hbar^3\pi^2}\sqrt{E}
$$
is density of states for the free electron gas and
\be
g(E)=\ln{{\bigg|}\frac{\sqrt{E_\textrm{F}/E}+1}
{\sqrt{E_\textrm{F}/E}-1}{\bigg|}-2\sqrt{E_\textrm{F}/E}}\,.
\ee

In order to calculate the averaged density of localized states, we use the relation
\be
\lim_{\varepsilon\rightarrow 0}\sum\limits_{\bf
k}\frac{|\Omega_k^\sigma|^2}{(E-E_k-\Lambda_0+\ri\varepsilon)^2}=
-\frac{\rd\Lambda^\sigma(E)}{\rd E}+\ri\frac{\rd\Delta^\sigma(E)}{\rd E}.
\label{quadrsum}
\ee
In the similar manner,
\begin{align}
\suml_{\bf k,q}&\frac{\Omega_k^\sigma{\Omega_q^*}^\sigma\ls n_{\bf
k-q}v({\bf |k-q|})+\,\widetilde v_0({\bf
|k-q|})\rs}{(E-E_k-\Lambda_0)(E-E_q-\Lambda_0)}=\nonumber\\
%\]
%\[
&={\cal P}\suml_{\bf k}\frac{|\Omega_k^\sigma|^2\Lambda^\sigma(k,\/E)}{E-E_k-\Lambda_0}-
\ri\pi\suml_{\bf k}|\Omega_k^\sigma|^2\Lambda^\sigma(k,\/E)\delta(E-E_k-\Lambda_0)\nonumber\\
%\]
%\be
&-\ri{\cal P}\suml_{\bf k}\frac{|\Omega_k^\sigma|^2\Delta^\sigma(k,\/E)}{E-E_k-\Lambda_0}-
\pi\suml_{\bf k}|\Omega_k^\sigma|^2\Delta^\sigma(k,\/E)\delta(E-E_k-\Lambda_0),
\end{align}
where we denote
\begin{align}
\Lambda^\sigma(k,E)&={\cal P}\suml_{\bf q}
\frac{{\Omega_q^*}^\sigma}{E-E_q-\Lambda_0} f(|\,{\bf k-\bf q}|\,), \\
%\ee
%\be
\Delta^\sigma(k,E)&=\pi\suml_{\bf q}{\Omega_q^*}^\sigma f(|\,{\bf k-\bf q}|\,)\delta(E-E_q-\Lambda_0),\\
%\ee
%\be
f(|\,{\bf k-\bf q}|\,)&=\big[n_{\bf k-q}v(|\,{\bf k-q}|\,)+
\widetilde v_0({|\,\bf k-q|}\,)\big].
\end{align}
The configuration averaged Green function has the form,
\be
\overline{L_{0,0}^\sigma}=\frac{1}{E-E_{0,\sigma}-\widetilde
\Lambda^\sigma(E)+\ri\widetilde \Delta^\sigma(E)}\,,
\ee
here,
\begin{align}
\widetilde \Lambda^\sigma(E)&=\Lambda^\sigma(E){+}\frac{\rd\Lambda^\sigma(E)}{\rd E}f(0)
+{\cal P}\suml_{\bf k}\frac{|\Omega_k^\sigma|^2\Lambda^\sigma(k,E)}{E-E_k-\Lambda_0}
-\pi\suml_{\bf k}|\Omega_k^\sigma|^2\Delta^\sigma(k,E)\delta(E-E_k-\Lambda_0),
\label{Lambda}\\
%\ee
%\be
\widetilde \Delta^\sigma(E)&=\Delta^\sigma(E){+}\frac{\rd\Delta^\sigma(E)}{\rd E}f(0)
+{\cal P}\suml_{\bf k}\frac{|\Omega_k^\sigma|^2\Delta^\sigma(k,E)}{E-E_k-\Lambda_0}
+\pi\suml_{\bf k}|\Omega_k^\sigma|^2\Lambda^\sigma(k,E)\delta(E-E_k-\Lambda_0)
\label{Delta}
\end{align}
are the effective shift and broadening of localized impurity level now contain
the structural disorder contribution, besides the contribution from interactions.

The occupation number of electrons for absolute zero temperature is
\be
\ang{n_{0\sigma}}=\ang{d_{0\sigma}^+d_{0\sigma}}=
\int\limits_{-\infty}^{E_\textrm{F}}\overline{\rho_0^\sigma(E)}\,\rd E,
\label{set_eq}
\ee
where
\be
\overline{\rho_0^\sigma(E)}=-\frac{1}{\pi}{\rm Im}
\,\overline{L_{0,0}^\sigma(E+\ri\varepsilon)}, \qquad  \varepsilon \rightarrow 0
\label{4.12}
\ee
is configurational density of localized states with the spin $\sigma$.
\be
\overline{\rho_0^\sigma(E)}=\frac{1}{\pi}\frac{\widetilde \Delta^\sigma(E)}{[E-E_{0,\sigma}-\widetilde\Lambda^\sigma(E)]^2+
[\widetilde \Delta^\sigma(E)]^2}\,.
\ee
After simple transformation  of the system of equations (\ref{Lambda})--(\ref{Delta}) we obtain:
\begin{align*}
\widetilde \Delta^\sigma(E)&=\pi[(\Omega^\sigma)^2+2\widetilde F \rho_0(E)+
g(E)]\rho_0(E)+
\pi (\Omega^\sigma)^2 f(0)\frac{\rd\rho_0(E)}{\rd E},\\
%\]
%\[
\widetilde\Lambda^\sigma(E)&=(\Omega^\sigma)^2 g(E)\rho_0(E)+
(\Omega^\sigma)^2 f(0)\frac{\rd g(E)\rho_0(E)}{\rd E}+
\widetilde F
{\rho_0(E)g(E)}^2{-}\pi^2\widetilde F \rho_0(E)^2.
\end{align*}
Here, the notation
\[
\ang{F} =\frac{\sum_{\bf k,\bf
q}\Omega_k^\sigma\Omega_q^\sigma f(|\,{\bf k-\bf
q}|\,)\delta(E{-}E_q{-}\Lambda_0)\delta(E{-}E_k{-}\Lambda_0)}
{1/V^2\sum_{\bf k,\bf
q}\delta(E{-}E_q{-}\Lambda_0)\delta(E{-}E_k{-}\Lambda_0)}
\]
is introduced for the average value of matrix elements $\Omega_k^\sigma\Omega_q^\sigma f(|\,{\bf k-\bf
q}|\,)$ at the Fermi level.

%%%%%%%%%%%%%%%%%%%%%%%%%%%%%%%%%%%%%%%%%%%%%%%

\setcounter{equation}{0}
\section{Results and discussions}

We need to calculate the average values of
matrix elements $\Omega_k^\sigma$ by using the formfactors of
scattering potentials (\ref{formfactors}). Ashcroft,
Heine-Aba\-ren\-kov, Cohen, Animalu model potentials are widely
applicable in liquid metal physics. The parameters of these potentials
are investigated and approved sufficiently completely, see e.g.
\cite{ostr,harr,Yuhn,Ashcroft-Lekner}. We have used the
Ashcroft's potential (including screening by the conduction electrons)
for the liquid sodium~\cite{ostr}.
%\be
%v(r)=\bigg\{\begin{array}{cc}
%  0 & r \leq r_\textrm{c}, \\
%  -Ze^2/r & r>r_\textrm{c}.
%\end{array}
%\label{v}
%\ee
The Fourier-transform of Ashcroft's potential is
\be
v(q)=-\frac{4 \pi Z e^2}{\Omega q^2}\cos(q r_\textrm{c}),
\label{v-q}
\ee
\noindent
where $r_\textrm{c}$ is the core radius.
The parameters for liquid sodium are $r_\textrm{c}^\textrm{Na}{=}0.0878$~nm,
$\Omega=270$~a.u.~--- atomic volume of liquid Na at $100~^\circ$C.
They are taken from the experimental data of resistivity measurements~\cite{ostr}.

The screened function by the conduction electrons in Heldart-Vosko
approximation is as follows:
\begin{align}
\varepsilon(q)&=1+\frac{4 \pi Z}{\Omega q^2}\lp\frac{2}{3}
E_\textrm{F}\rp^{-1}\lambda\lp\frac{q}{2k_\textrm{F}}\rp \ls 1-f(q)\rs,\nonumber\\
%\]
%\[
 \lambda(y)&=\frac{1}{2}+\frac{1-y^2}{4 y}\ln \bigg|\frac{1+y}{1-y}\bigg|,\nonumber\\
%\]
%\be
f(q)&=\frac{1/2q^2}{q^2+2k_\textrm{F}/(1+0.01574(\Omega/Z)^{1/3})},
\label{epsilon-q}
\end{align}
where  $k_\textrm{F}=(3 \pi^2 Z/\Omega)^{1/3}=0.4786$~a.u.$^{-1}$.

Now, let us consider the interaction between the electron and the
negative ion (\ref{pseudopot}).
Different forms of polarization potential were discussed in~\cite{Smirnow,Golov,Cooper,Robinson}.

We have proposed a new model potential for electron-negative ion
interaction in \cite{modpsepot}:
\[
V_0(r)=A\frac{\re^{-r/r_\textrm{p}}}{r^2}-\frac{\alpha}{(r^2+r_\textrm{p}^2)^2}\,.
\]
where $A=3/8r_\textrm{p}^2E_0+3\alpha I/r_\textrm{p}^2-3/8$.
The semi-empirical parameters
$\alpha$ and $r_\textrm{p}$ do not arise naturally from the formalism. Thus,
the only criterium available to establish the accuracy of the method is in the
agreement with the experimental results.
Hence, we use the values of $r_\textrm{p}$ \cite{Cooper} and $\alpha$ \cite{Robinson}
taken from the experimental data for electron photodetachment from negative ions.

The formfactor of the effective impurity potential in Hartree-Fock approximation is
\be
\widetilde v_0(q)=\frac{8\pi A}{q} {\arctan} (qr_\textrm{p})-
\frac{4\alpha \pi^2}{r_\textrm{p}} \re^{-qr_\textrm{p}}+
\ang{\widehat n}\frac{8\pi}{q^2}\bigg[1-\frac{q^2 r_\textrm{p}^2}{(4+q^2r_\textrm{p}^2)}
\bigg(1+\frac{4}{4+q^2r_\textrm{p}^2}\bigg)\bigg].
\ee

The correlation function is as follows:
\[
n_q=1+\frac{3\eta}{(qr_\textrm{c})^3\left[qr^*\cos(qr^*)-\sin(qr^*)\right]}\,,
\]
where $r^*=r_\textrm{c}+r_\textrm{p}$.

By using the expressions for the model potentials of liquid metal and impurity
and for the correlation function, one can calculate the average value of $\Omega^\sigma$, $\ang{F}$.

The potentials $w, u$ were discussed in the work~\cite{JPS}, specifically at Fermi level
$(w)^2 \approx E^2/(E\gamma+1)^4$, $\gamma=2mr_\textrm{p}^2/\hbar^2$, and
$z=u/w$ is the parameter of intensity of scattering process on the charged impurity.
The function $K(k,q)$ on the angles in a spherical coordinate system, was introduced to simplify the calculation of
$\ang{F}$:
\[
K(k,q)=2\pi\intl_0^\pi \sin\theta
f\lp\sqrt{k^2+q^2-2kq \cos\theta}\rp \rd\theta.
\]
\noindent
Then, using $\sum_{\bf k}{=}[V/(2\pi)^3]\int_0^\infty k^2 \rd k \int_0^{2\pi} \rd\varphi$,
 we obtain the averaged value $\ang{F}$ that characterizes the structural contribution.

The parameter $\delta = \ang{F} \rho_0 g
/{(\Omega)}^2$ measures the value of disorder, and  we assume  $0<\delta
\ll 1$. The parameter $h=\Lambda^\sigma/\Delta^\sigma$ has the meaning of a local level shift.

\begin{figure}[!b]
\centerline{
\includegraphics[width=0.42\textwidth]{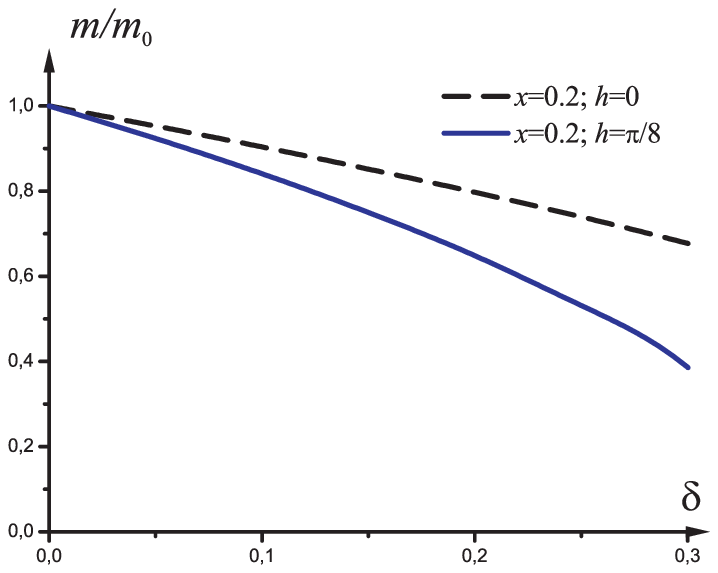}
\hspace{10mm}
\includegraphics[width=0.45\textwidth]{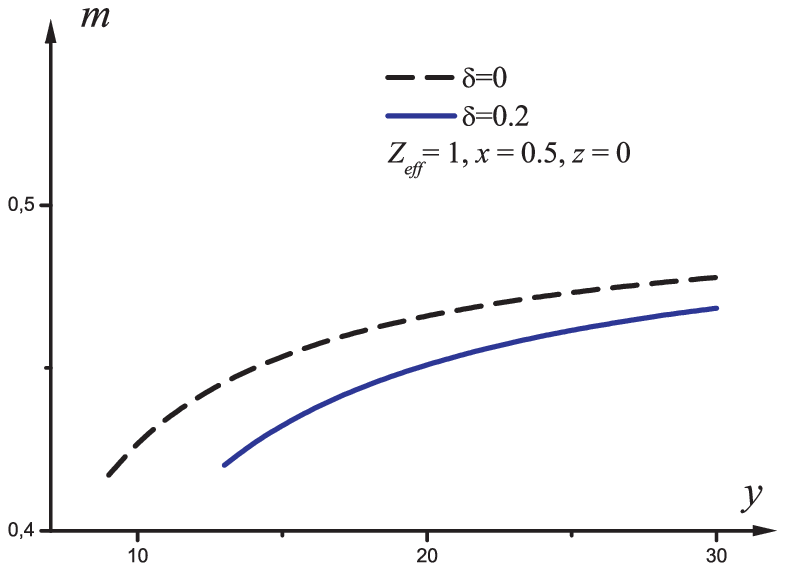}
}
\parbox[t]{0.5\textwidth}{
\caption{(Color online) The dependence of the impurity magnetic moment on the structural parameter
$\delta$ ($y=10$, $x=0.2$, $z=0.1$, $h=\pi/8$).}\label{mx02}
}
\parbox[t]{0.5\textwidth}{
\caption{(Color online) The dependence of the impurity magnetic moment on the degree of Coulomb repulsion $y=U_0/\Delta$
at constant $Z_\textrm{eff}=1$.} \label{m_y}
}
\end{figure}

The dimensionless value  $x=(E_\textrm{F}-E_0)/U_0$ means
that the local impurity level lies on the Fermi level for $x=0$.
For $x=1$, the Fermi level lies on $E_0+U_0$. For magnetic solutions
$x=1/2$, which means, of course, that the Fermi level is
exactly halfway between the case where only one electron is in a localized state
and the one in which both are in the same state with opposite spins.
The parameter $y=U_0/\Delta$ measures the ratio of Coulomb integral respective to the width
of virtual state.

The spin-polarized magnetic impurity state $m=\ang{n_{+}-n_{-}}$, $m/m_0$
is shown in figure~\ref{mx02}. Here, $m_0=0.849 \mu_\textrm{B}$
for $\delta=0$ (the quasi-crystalline case). The decrease of impurity local
magnetic moment with the growing $\delta$ is shown in figure~\ref{mx02}.
The additional local level shift due to the interaction of condition electrons with the impurity
leads to a decrease of the local magnetic moment.

\begin{figure}[!t]
\centerline{
\includegraphics[width=0.55\textwidth]{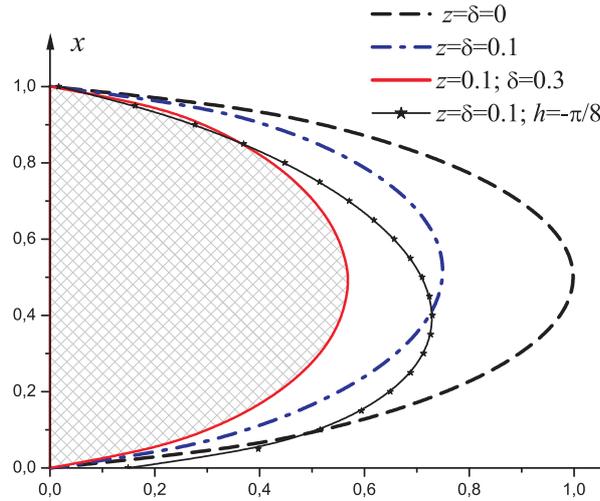}
}
\caption{(Color online) The phase diagram exhibits the regions with magnetic and nonmagnetic states.}
\label{fazdiagr}
\end{figure}

The  behavior of magnetic moment at constant value of the effective impurity charge
is presented in figure~\ref{m_y} when the parameter $y=U_0/\Delta$ increases. When $y$ is large but finite, magnetic solutions are still possible but as $y$ is reduced they eventually disappear.

The diagram describing the region of existence of magnetic and nonmagnetic states is
presented in figure~\ref{fazdiagr}. The interplay of hybridization and local environment disorder produces a rich structure zero-temperature phase diagram.
The region of impurity magnetic states in a disordered metal decreases in contrast to
the quasi-crystalline case ($\delta=0$) \cite{prepr0619}. This famous experimental fact for ferromagnetic alloys is discussed in various monographs, see e.g.~\cite{xandrix}.

\begin{figure}[!t]
\centerline{
\includegraphics[width=0.6\textwidth]{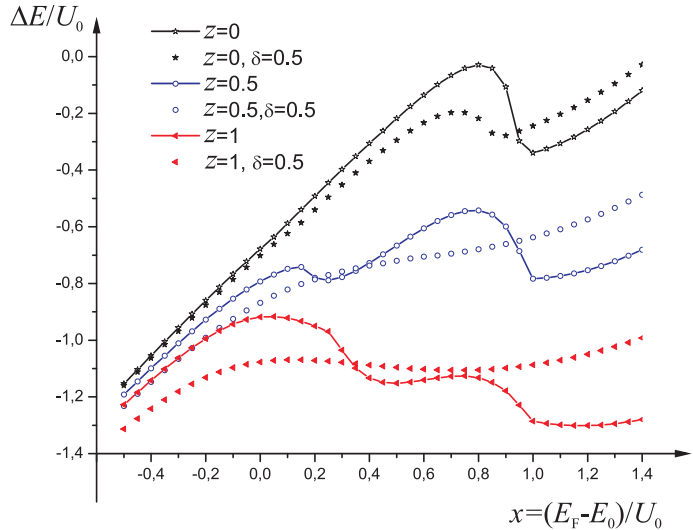}
}
\caption{(Color online) The solvation free energy of impurity atom in liquid metal.} \label{solv}
\end{figure}

The solvation free energy, $\Delta E$, that determines the
excess free energy associated with the insertion of an impurity
atom into liquid metal, was calculated for this model in
our earlier work~\cite{UJP} that corresponds to the quasi-crystalline case.
The dependence of $\Delta E$, caused by the impurity solvation in liquid metal,
on Fermi level $x$, is shown in figure~(\ref{solv}).
The dotted lines correspond to the cases when the structural disorder is taken into account.
The solid lines correspond to the quasi-crystalline approximation~\cite{UJP}.

\section{Conclusions}

A generalized model proposed in this article permits to calculate the  microscopic
characteristics of impurity states in liquid  metal and to analyze the effect of
the structural disorder on the macroscopic properties.

Using the equation of motion method for the two-time retarded Green function and
using HFA, the system of self-consistent equations for average
thermodynamic occupation numbers of localized impurity level is obtained.
The region of impurity magnetic states in a disordered metal decreases in contrast to the
quasi-crystalline case. The contribution to the broadening of virtual impurity level
at $T=0$ comes from the scattering processes on the charged impurity and from the structural disorder of the impurity environment as well.  This interplay may be relevant to experimental realizations of
the system ``liquid metal+electronegative impurity'' in order to study its magnetic properties.

The next possible step of exploration of the proposed model can be the study of Kondo regime
taking into account the processes of exchange.
In the discussed Hamiltonian (\ref{Hamiltonian}), these processes are described by the following terms,
$a_{\bf{k}1\sigma}^+\,a_{\bf{k}2\sigma'}^+d_{0,\sigma'}d_{0,\sigma}$, \, $d_{0,\sigma'}^+d_{0,\sigma'}^+a_{\bf{k}2\sigma'}\,a_{\bf{k}1\sigma}$, \,
$a_{\bf{k}1\sigma}^+\,d_{0,\sigma'}^+a_{\bf{k}2\sigma'}d_{0,\sigma}$, that
correspond to the spin flip processes. They were not accounted for because
their matrix elements are of an order of magnitude less than the Coulomb matrix elements.
However, using these terms and the decoupling scheme beyond the HFA one can analyse
the Kondo effect, which is important at low temperatures. This exchange interaction
is more likely to increase the polarization of the band electrons rather than to enhance the formation of a magnetic moment.

%%%%%%%%%%%%%%%%%%%%%%%%%%%%%%%%%%

%\newpage

\ukrainianpart
\title{Магнітні стани ізольованої домішки у невпорядкованому середовищі}

\author{Г.В.~Понеділок, М.І.~Клапчук}
\address{Національний університет ''Львівська політехніка'', Інститут прикладної математики \\
та фундаментальних наук, вул.~С.~Бандери, 12, 79013 Львів, Україна}

\makeukrtitle

\begin{abstract}
\tolerance=3000%
Досліджується зарядовий та магнітний стани домішки,
розчиненої  в аморфному металічному сплаві. Гамільтоніан системи є
узагальненням моделі Андерсона, де додатково враховано процеси
пружнього  і непружнього розсіяння електронів провідності на іонах
металу та на зарядженій домішці. Пропонується метод розрахунку
конфігураційно усереднених одноелектронних функцій Гріна в
наближенні Хартрі-Фока. Отримана система   самоузгоджених рівнянь
для розрахунку зарядового та спін-поляризованого стану домішки.
Подано якісний аналіз впливу структурної невпорядкованості металевої
матриці на спостережувані величини. Показано, що структурний безлад
середовища приводить до додаткового розширення та зсуву віртуального
енергетичного рівня домішки, зменшуючи магнітний момент домішки.
\keywords домішкові стани, структурний безлад

\end{abstract}

\end{document}